\documentclass[amsmath,amssymb,twocolumn,aps]{revtex4-1}         

\usepackage{amssymb}
\usepackage{multirow}
\usepackage{amsfonts}
\usepackage{amsmath}
\usepackage{amssymb}
\usepackage{graphicx}
\usepackage{dcolumn}
\usepackage{times}
\usepackage{color}

\begin{document}

\title{Symmetric solitonic excitations of the (1+1)-dimensional Abelian-Higgs ``classical 
vacuum'' }

\author{F.K. Diakonos$^{1}$}
%\email[]{fdiakono@phys.uoa.gr}
\author{G.C. Katsimiga$^{1}$}
\author{X.N. Maintas$^{1}$}
\author{C.E. Tsagkarakis$^{1}$}
%\email[]{xtsagkarakis@gmail.com}

\affiliation{$^{1}$Department of Physics, University of Athens, GR-15784 Athens, Greece}
%\affiliation{$^{2}$Department of Physics, Florida State University, Tallahassee,
%Florida, USA}
%\affiliation{$^{3}$Hellenic Naval Academy, Hatzikyriakou Avenue, Pireaus 185 39, Greece}
\date{\today}

\begin{abstract}

We study the classical dynamics of the Abelian-Higgs model in (1+1) space-time dimensions 
for the  case of  strongly broken gauge symmetry. 
In this limit the wells of the potential are almost harmonic and sufficiently deep,  
presenting a scenario far from the associated critical point. Using a multiscale perturbation 
expansion, the equations of motion for the fields are  reduced to a system of coupled nonlinear 
Schr\"{o}dinger equations (CNLS). Exact solutions of the latter are used to obtain approximate 
analytical solutions for the full dynamics of both the gauge and Higgs field in the form of 
oscillons and oscillating kinks. Numerical simulations of the exact dynamics verify the validity 
of these solutions. We  explore their persistence for a wide range of the model's single 
parameter which is the ratio of the Higgs  mass $(m_H)$ to the gauge field mass $(m_A)$. We show 
that only oscillons oscillating symmetrically with respect to the ``classical vacuum'', for both 
the gauge and the Higgs field,
are long lived. Furthermore plane waves and oscillating kinks are shown to decay into oscillon-
like patterns, 
%dominating the solution space of the model 
due to the modulation instability mechanism. 

\end{abstract}

\maketitle

%%%%%%%%%%%%%%%%%%%%%%%%%%
  
\section{Introduction}

The study of solitons in nonlinear field theories, still continues to attract considerable 
attention. This is due to the fact that such field configurations  are relevant for the phenomenological description of a wide class of physical systems ranging from elementary particles to superconductors and Bose-Einstein condesates. 
Being localized structures characterized by a coherent time evolution, solitons can be related 
to particle like structures such as, 
magnetic monopoles, sphalerons, extended structures in the form of domain walls and 
cosmic strings, and non-topological localized structures namely oscillons, having implications 
for the cosmology of the early 
universe~\cite{earlyuniverse,amin,stama068,khlopov,
hybrid,stama068,glegra2014,amin2010,amin12,linde,copeland2,glegrastam}. 

 Usually localized solitonic states emerge as solutions in field theories  characterized by a 
scale which may originate from the spontaneous breaking of an underlying symmetry. The simplest 
models in which such a scenario can be realized are  scalar theories ~\cite{rajamaran, 
manton,neveu,bogo,fodor09,hertzberg2010,campbell,salmi08,
gleiser08,saffin07,copeland,honda,fodor08,amin2} admitting a variety of such solutions like the 
exact kink solution~\cite{manton} or the breather~\cite{segur} of the $\phi^4$ theory. 
Furthermore, models with scalars coupled to gauge fields, also admit various types of 
soliton solutions  that can be obtained both analytically and 
numerically~\cite{arnold,prd,gleiser21,farhi05,graham,sfakia}.

 Among the models involving a gauge field coupled with a scalar, the Abelian- Higgs model in 
$(1+1)$ dimensions shares distiguinshed ground. Firstly it may reveal important features of 
superconductivity~\cite{weinberg,gl} allowing at the same time for analytical treatment. In fact 
the search for one dimensional solutions of the effective theory of 
superconductivity~\cite{1sc}, namely the Gor’kov-Eliashberg-Landau-Ginzburg approach, can be 
directly associated with the investigation of classical solutions of the (1+1)-dimensional 
Abelian-Higgs model~\cite{prd}.
Additionally, in most recent studies this model plays an important role in one dimensional
holographic superconductors~\cite{holo}. Finally the Abelian-Higgs model in (1+1)-dimensions has 
been used as a toy model for the 
description of topological charge fluctuations of the vacuum at zero temperature via instantons 
or the study of sphalerons at finite temperature. Both of these field configurations 
(instantons, sphalerons), have been associated in previous studies with the baryon number 
violation in the Universe~\cite{ukawa,rubakov}.

Here we are interested in studying  soliton solutions of the (1+1)-dimensional Abelian-
Higgs model, focusing on oscillons and oscillating kinks. In particular oscillons, also known as 
breathers, are of special interest since they can be linked to the dynamics of the early 
Universe~\cite{amin12} and to the Meissner effect of superconductivity~\cite{prd}. 
In most cases oscillons are found numerically, however 
approximate analytical solutions describing oscillons were rigorously derived in~\cite{prd} by 
introducing multiple scale perturbation expansion~\cite{multiplescales} in gauge 
theories~\cite{emeis}, and assuming a sufficiently small amplitude for the Higgs field. In this 
limit the dynamics simplify considerable and it is found that the scalar field  performs 
asymmetric oscillations around the ``classical vacuum''. Such a scenario occurs naturally 
considering the model just after the symmetry breaking i.e. close to the critical point. Then 
the minimum of the scalar field potential is very flat and  the asymmetric cubic term is strong 
leading to an asymmetric shape of the potential around it.

In the present work we investigate the (1+1)-dimensional Abelian-Higgs model in a different 
limit, where the gauge and the scalar field amplitudes are of the same order. 
This scenario corresponds to a strong 
breaking of the underlying gauge symmetry, far beyond the related critical point. In this case 
the minimum of the potential occurs at the 
bottom of a deep well, while the potential shape is almost symmetric around it since the 
quadratic term dominates. The resulting 
dynamics, derived within the framework of multiple scale perturbation theory 
are significantly more complex leading to a system of  CNLS equations describing the field 
envelopes. 
Solving analytically the coupled system we obtain bright-bright, dark-bright and 
dark-dark soliton solutions. Our analysis shows that only bright-bright solitons may describe 
long-lived oscillons of the original equations of motion, with both fields  performing  
symmetric oscillations around the ``classical vacuum''. Additionally, it is shown that any dark 
component is subject to the modulation instability (MI) mechanism~\cite{mi}.
Subsequently, we integrate numerically 
the exact equations of motion and we verify that the analytically
obtained solutions describe the long-lived oscillons sufficiently well.
Surprisingly enough, even when the perturbation expansion is not valid, 
using  initial conditions corresponding to bright-bright solitons, we numerically obtain  
long-lived oscillons. Furthermore, we use the MI mechanism in order to illustrate
that oscillons may be formed for all values of the parameter $m_H/m_A$. 

The paper is organized as follows: in section II, we present 
the equations of motion for the Abelian-Higgs model   
reducing them to a CNLS system with the method of multiple scales and provide  
the corresponding analytical solutions.
In section III we present results of direct numerical integration of the original equations of
motion to check the validity of our approximations and study the stability of the 
analytically found solutions.
Finally, in section IV we present our concluding remarks.
\vspace{-10pt}
%%%%%%%%%%%%%%%%
\section{General formalism and setup}

\subsection{Deriving the NLS equations}

The Lagrangian density of the model has the form:
%%%%
\begin{eqnarray}
{\mathcal{L}}=-\frac{1}{4} F_{\mu \nu} F^{ \mu \nu} &+& (D_\mu \Phi)^{*}(D^\mu \Phi)
- V(\Phi^{*} \Phi),
\label{l}
\end{eqnarray}
%%%%
where $\Phi$ is a complex scalar field, $D_\mu=\partial_\mu + ieA_\mu$ is the
covariant derivative
with $e$ the coupling constant, and $F_{\mu \nu}$ is the electromagnetic tensor. The potential
$V(\Phi^{*} \Phi)$
has the form:
%%%
\begin{eqnarray}
V(\Phi^{*} \Phi)=\mu^2 \Phi^{*} \Phi + \lambda (\Phi^{*} \Phi)^2,
\label{V}
\end{eqnarray}
%%%%
with $\mu^2<0$ and $\lambda > 0$ being undefined constants. 
For the spontaneously broken symmetry case, we choose as vacuum
the minimum $\upsilon=\sqrt{-\mu^2/2\lambda}$ of the potential given by Eq.~(\ref{V}).
We expand the $\Phi$
field around this vacuum expectation value (vev) as $\Phi=\upsilon+ H/ \sqrt{2}$, gauging
away the Goldstone mode. $H$ is a real scalar field, the Higgs field, with mass 
$m_H=\sqrt{2\lambda\upsilon^2}$.
Due to the symmetry breaking the gauge field $A_{\mu}$ acquires mass $m_A=e\upsilon$. 

%We consider the
%ansatz: $A_0=A_1=A_3=0$ and $A_2=A(x,t)$ which is compatible with the Lorentz condition
%in $(1+1)$-dim and simplifies significantly the equations of motion.  
We reduce the theory to a $(1+1)$-dimensional model by considering the ansatz: $A_0=A_1=A_3=0$ 
and $A_2=A(x,t)$ which is compatible with the Lorentz condition and simplifies significantly the 
equations of motion.
Defining dimensionless variables: $\tilde{x}^{\mu}=m_Ax^{\mu}$,  
$\tilde{A}=(e/m_A)A$, $\tilde{H}=(e/m_A)H$ 
[vev is also scaled as $\tilde{\upsilon}=(e/m_A)\upsilon$], and 
dropping the tildes after this substitution, the corresponding equations of motion become:
%%%
\begin{eqnarray}
(\Box &+& 1) A + 2 H A +  H^2 A = 0,
\label{eqa}\\
(\Box &+& q^2 ) H + \frac{3}{2} q^2 H^2 + \frac{q^2}{2} H^3 +  A^2 (1 + H) = 0,
%\nonumber \\
\label{eqh}
\end{eqnarray}
%%%
where $A,H$ are functions of $(x,t)$, while $q \equiv m_H/m_A$ is the single dimensionless 
parameter that designates the dynamics.
The energy density corresponding to the above equations of motion is
\begin{equation}
E(x,t) = \frac{1}{2}\left[\left(\partial_t A\right)^2+\left(\partial_x A\right)^2+\left(\partial_t H\right)^2+\left(\partial_x H\right)^2 \right]+V,
\label{energy}
\end{equation}
with $V=q^2H^2\left(H+2\right)^2/8+A^2\left(H+1\right)^2/2$, being the potential energy.
We are interested in finding localized solutions to the above system of equations i.e.
Eqs.~(\ref{eqa})-(\ref{eqh}). For this purpose we 
employ multiple scale perturbation theory~\cite{multiplescales} expanding space-time 
coordinates
and their derivatives as follows: $x_0=x$, $x_1=\epsilon x$, $x_2=\epsilon^2 x, \ldots$,
$t_0=t$, $t_1=\epsilon t$, $t_2=\epsilon^2 t, \ldots$, 
$\partial_x = \partial_{x_0} + \epsilon \partial_{x_1} + \ldots$,
$\partial_t = \partial_{t_0} + \epsilon \partial_{t_1} + \ldots$. Accordingly we write the
gauge and the scalar field as:
%%%
\begin{eqnarray}
A&=& \epsilon A^{(1)} + \epsilon^2 A^{(2)} + \ldots , \label{expA}
\\
H&=& \epsilon H^{(1)} + \epsilon^2 H^{(2)}+ \ldots ,  \label{exph}
\end{eqnarray}
%%%%
where $\epsilon$ is a formal small parameter: $0<\epsilon\ll1$, related to the
amplitude of the Higgs and the gauge field excitations around the physical vacuum.
Inserting  Eqs.~(\ref{expA})-(\ref{exph}) into Eqs.~(\ref{eqa})-(\ref{eqh}), and
expressing the operators in
terms of the slow scales mentioned above, we proceed in our analysis 
solving the equations of motion order by order.
To first order in $\epsilon$ we
have the following decoupled equations for the gauge (A) and the Higgs (H) field respectively:
%%%%
\begin{eqnarray}
\!\!\!\!\!\!\!\!\!\!\!\mathcal{O}(\epsilon)\!\!:&(\Box_0 + 1)&A^{(1)}=0,
 \label{order1a}
 \\
\!\!\!\!\!\!\!\!\!\!\!\!\!\!\!\!\!\!\!\!\!\!\!\!\!\!&(\Box_0 + q^2)&H^{(1)}=0.
 \label{order1h}
\end{eqnarray}
%%%%
Equations (\ref{order1a}) and (\ref{order1h}), acquire plane wave
solutions of the form:
%%%%
\begin{eqnarray}
 A^{(1)}&=&fe^{i\theta_1}+f^{*} e^{-i\theta_1},
 \label{sol1a}
 \\
 H^{(1)}&=&le^{i\theta_2}+l^{*} e^{-i\theta_2},
 \label{sol1h}
\end{eqnarray}
%%%%
where ``*'' denotes complex conjugate, while $f=f(x_i,t_i)$ and $l=l(x_i,t_i)$ are functions
of the slow variables that have to be determined (the index $i=1,2,\ldots$ refers 
to the slow scales).
The phase $\theta_j$ is defined as: $\theta_j \equiv k_j x-\omega_j t$,
where the index $j=1,2$ refers to the gauge and the scalar
field respectively.
Substituting Eqs.~(\ref{sol1a}) and (\ref{sol1h}) into Eqs.~(\ref{order1a})-(\ref{order1h}) we 
get the dispersion relations for the two fields i.e. $\omega^2_1=k^2_1+1$ and
$\omega^2_2=k^2_2+q^2$.
Thus, to first order in $\epsilon$ the linear limit of the theory is recovered.
Proceeding to the next order of the perturbation scheme, namely $\mathcal{O}(\epsilon^2)$, 
we obtain the following system
of equations:
%%%%%
\begin{eqnarray}
\left(\Box_0 +1\right)A^{(2)}&=&-2\partial_{\mu_0} \partial^{\mu_1}A^{(1)}-2H^{(1)}A^{(1)},
 \label{order2a} \\
(\Box_0 + q^2)H^{(2)}&=&-2\partial_{\mu_0} \partial^{\mu_1}H^{(1)}-\left(\frac{3q^2}{2} H^{(1)2}
+
A^{(1)2}\right). \nonumber \\
\label{order2h}
\end{eqnarray}
%%%
Notice that the first terms on the right hand side of the above equations, are ``$secular$'',
that is in resonance with the operators on the left side. 
These terms imply a linear growing of $A^{(2)}$, $H^{(2)}$  
with time and therefore lead to the blow-up of the solutions.
Consequently, in order for the perturbation scheme to be valid, these terms have to vanish 
independently, leading to the following equations for  $f=f(x_i,t_i)$ and $l=l(x_i,t_i)$:
%%%%%%%%%%5
\begin{eqnarray}
\hat{L}_1f&=&0,
\label{seca} \\
\hat{L}_2l&=&0.
\label{sech}
\end{eqnarray}
%%%%%%%%%
The operator $\hat{L}_j$ is defined as
$\hat{L}_j\equiv-i\left(\partial_{t_1}+\upsilon^{(j)}_{g}\partial_{x_1}\right)$,
with $\upsilon^{(j)}_g\equiv\partial \omega_{j}(k)/\partial k_{j}=k_{j}/\omega_{j}$
being the group velocities for the
gauge (j=1) and the Higgs (j=2) field respectively.  
Eqs.~(\ref{seca})-(\ref{sech}) are automatically
satisfied if $f=f\left(X_1,t_2\right)$ and $l=l\left(X_2,t_2\right)$ where
$X_j\equiv x_1-\upsilon^{(j)}_g t_1$.
%\end{widetext}

Furthermore, Eqs.~(\ref{order2a})-(\ref{order2h}) can be solved analytically and the
solutions for the fields $A^{(2)}$ and $H^{(2)}$ as functions of $f$ and $l$ have the following
form:
%%%%
\begin{align}
&A^{(2)}=\frac{fl}{a}e^{i\Theta_{+}}+\frac{fl^{*}}{b} e^{i\Theta_{-}}+ c.c.,
\label{a2}\\
&H^{(2)}=\frac{l^2}{2}e^{2i\theta_{2}}+\frac{f^2}{4-q^2} e^{2i\theta_{1}}-\frac{3q^2|l|^2+2|f|
^2}{q^2}+ c.c.,
%\nonumber\\
\label{h2}	
\end{align}
%%%
where ``c.c.'', stands for the complex conjugate, while
$a=\omega_1\omega_2-k_1k_2+q^2/2$,
$b=k_1k_2-\omega_1\omega_2+q^2/2$ and
$\Theta_{\pm}\equiv \theta_1 \pm \theta_2$.
Note, that the coefficients in equations~(\ref{a2})-(\ref{h2}), e.g. $1/b$,
define regions for the parameter $q$ for which
the fields $A^{(2)}$ and $H^{(2)}$ could become infinitely large, 
[in the perturbation are assumed to be of $\mathcal{O}(1)$].
Such regions are for consistency excluded in our analysis.
Also notice that once the functions $f$ and $l$ are determined, the solutions for $A^{(2)}$ and
$H^{(2)}$ are also fixed.
Continuing our analysis we get to $\mathcal{O}(\epsilon^3)$ the following equations:
%%%%
\begin{align}
\left(\Box_0 + 1\right)A^{(3)}=-2\partial_{\mu_0}\partial^{\mu_1}A^{(2)}-
\left(\Box_1+2\partial_{\mu_0}\partial^{\mu_2}\right)A^{(1)} \nonumber \\
-2\left(H^{(2)}A^{(1)}+A^{(2)}H^{(1)}\right)-H^{(1)2}A^{(1)} , %\nonumber \\
\label{order3a} \\
\left(\Box_0 + q^2\right)H^{(3)}=-2\partial_{\mu_0}\partial^{\mu_1}H^{(2)}-
\left(\Box_1+2\partial_{\mu_0}\partial^{\mu_2}\right)H^{(1)} \nonumber \\
-3q^2H^{(1)}H^{(2)}-\frac{q^2}{2}H^{(1)3}
-2A^{(1)}A^{(2)}-A^{(1)2}H^{(1)}.
\label{order3h}
\end{align}
%%%
The first terms on the right hand side of Eqs.~(\ref{order3a})-(\ref{order3h})
can be eliminated through the aforementioned choice for the variables $X_i$ 
along with the condition $\upsilon^{(1)}_{g}=\upsilon^{(2)}_{g}$.  
Furthermore, to simplify the calculations remaining consistent, we choose $k_1=k_2=0$, i.e.
$\upsilon^{(1)}_{g}=\upsilon^{(2)}_{g}=0$ so that $X_1=X_2=x_1$. 
Additionally, at the same order, the solvability condition 
requires the $secular$ parts of 
Eqs.~(\ref{order3a})-(\ref{order3h}) to vanish,
leading to the following equations:
\begin{align}
\left(\Box_1 +2\partial_{\mu_0}\partial^{\mu_2}+2H^{(2)}+H^{(1)2}\right)A^{(1)}
&=-2A^{(2)}H^{(1)},
\label{comp1} \\
\Big(\Box_1 +2\partial_{\mu_0}\partial^{\mu_2}+3q^2H^{(2)}+\frac{q^2}
{2}H^{(1)2}&+A^{(1)2}\Big)H^{(1)} \nonumber \\
&= -2A^{(1)}A^{(2)}. 
\label{comp2}
\end{align}
With the above assumptions we derive a system of
CNLS equations for the functions $f(x_1,t_2)$ and $l(x_1,t_2)$:
%%%%
\begin{eqnarray}
i\partial_{t_2}f&=&-\frac{1}{2}\partial^{2}_{x_1}f+g_{11}|f|^2f+g_{12}|l|^2f ,
\label{nlsf} \\
iq\partial_{t_2}l&=&-\frac{1}{2}\partial^{2}_{x_1}l+g_{21}|f|^2l+g_{22}|l|^2l ,
\label{nlsl}
\end{eqnarray}
%%%%
where $g_{ij} \equiv g_{ij}(q)$ are the following functions of $q$:
%%%
\begin{eqnarray}
g_{11}&=&-\left(\frac{2}{q^2}+\frac{1}{q^2-4}\right), \quad \!\!\!\!g_{12}=-\left(2-\frac{4}
{q^2-4}\right)\nonumber \\
g_{21}&=&g_{12}, \quad \;\;\;\;\;\;\;\;\;\;\;\;\;\;\;\;\;\;\;\;\;\; g_{22}=-3q^2.
\label{gij}
\end{eqnarray}
%%%%

\subsection{Soliton solutions of the NLS equations and approximate oscillons
and oscillating kinks}
%%%%%
\begin{figure}[tbp]
\centering
\includegraphics[scale=0.4]{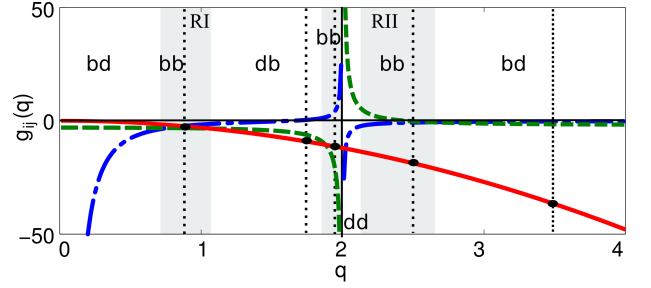}
\caption{(Color online): Sketch showing the regions of existence of each type of solution,
depending on different values of the parameter $q$. Dashed-dotted [blue] lines correspond to 
$g_{11}(q)$, solid [red] line
to $g_{22}(q)$, and dashed [green] lines to
$g_{12}(q)=g_{21}(q)$. Moreover vertical dashed [black] lines at $q=0.88$, $q=1.75$, $q=1.95$, 
$q=2.5$ and $q=3.5$ respectively, correspond to the parameters used for the simulations given 
below.}
\label{fig1}
\end{figure}
%%%%%%%%%%%%%%%%
%\begin{widetext}
%%%%%%
%\begin{center}
\begin{table*}[t]
\centering
\caption{Normalized amplitudes widths and frequencies}
\centering
\begin{tabular}{|c|c|c|c|c|}
\hline
R & $(a_2/a_{1})^2\equiv\bar{a}$ & $\left(\beta /a_{1} \right)^2\equiv\bar{\beta}$ & $\left(\nu_1/a^{2}_{1}\right)$  & $\left(\nu_{1}/\nu_{2}\right)$ 
 \\ [0.5ex]
\hline
 $bb$ & $\left(g_{11}-g_{21}\right)/\left(g_{22}-g_{12}\right)$ & $\left(g_{12}^2-
 g_{11}g_{22}\right)/\left(g_{22}-g_{11}\right)$ & $-\bar{\beta}/2$ & $q$ \\ [0.5ex] \hline
 $dd$ & $\left(g_{11}-g_{21}\right)/\left(g_{22}-g_{12}\right)$ & $\left(g_{11}g_{22}-
 g_{12}^2\right)/\left(g_{22}-g_{12}\right)$ & $\bar{\beta}$ & $q$ \\ [0.5ex] \hline
 $db$ & $\left(g_{11}-g_{21}\right)/\left(g_{12}-g_{22}\right)$  & $\left(g_{12}^2-
 g_{11}g_{22}\right)/\left(g_{12}-g_{22}\right)$ & $\bar{\beta}+g_{12}\bar{a}$ & \begin{tabular}{c}
$2qg_{11}\left(g_{12}-g_{22}\right)$\\ \hline $\left(g_{12}^2-2g_{12}g_{22}+g_{11}g_{22}\right)$ \end{tabular}
\\
[0.5ex] \hline
 $bd$ & $\left(g_{11}-g_{21}\right)/\left(g_{12}-g_{22}\right)$ & $\left(g_{11}g_{22}-
 g_{12}^2\right)/\left(g_{12}-g_{22}\right)$ & $-\bar{\beta}/2+g_{12}\bar{a}$ & \begin{tabular}{c} 
 $-2qg_{11}\left(g_{12}-g_{22}\right)$ \\ \hline $ \left(g_{12}^2-2g_{12}g_{22}+g_{11}g_{22}\right)$ 
\end{tabular}
\\
[0.3ex]
\hline
\end{tabular}
\label{table:sol}
\end{table*}
%\end{center}
%%%%%
The system of Eqs.~(\ref{nlsf})-(\ref{nlsl}), in the limit $g_{ij}=1$, is reduced to the
integrable Manakov model~\cite{manakov}, which admits exact analytical solutions.
For $g_{ij}\ne 1$ analytical soliton solutions of the  CNLS system can also be obtained, as it 
was shown in a recent work~\cite{veldes}.
In our case the coupling constants $g_{ij}$ depend on the parameter $q$ (cf. Fig.~\ref{fig1})
and in general $g_{ij}\ne 1$. 
Adopting the method of~\cite{veldes} we show below that solitons can be obtained 
in different $q$ regions.
First we look for solutions in the form of bright-bright ($bb$) solitons using the following
ansatz: 
%%%%
\begin{eqnarray}
f_{bb}=a_1 {\rm sech} \left(\beta_{bb}x_1\right)e^{-i\nu_1t_2},
\label{fbb} \\
l_{bb}=a_{2}{\rm sech}\left(\beta_{bb}x_1\right)e^{-i\nu_2t_2},
\label{lbb}
\end{eqnarray}
%%%
where $a_{1,2}$ denote the amplitudes, $\nu_{1,2}$ the frequencies of the solitons,
and $\beta_{bb}$ is the inverse width of the solitons. 
The condition that the squared amplitudes and widths of the solutions are positive defines
three regions of the single parameter $q$ where $bb$ solitons can be found, i.e.
$0.76<q<1.06$, $1.88<q<2$ and $2.13<q<2.63$.
Another type of solution, in the form of dark-dark ($dd$) solitons, can also be obtained using 
the ansatz
%%%%%
\begin{eqnarray}
f_{dd}=a_1\tanh\left(\beta_{dd}x_1\right)e^{-i\nu_1t_2},
\label{fdd} \\
l_{dd}=a_{2}\tanh\left(\beta_{dd}x_1\right)e^{-i\nu_2t_2}.
\label{ldd}
\end{eqnarray}
%%%%
The aforementioned condition for the amplitudes and widths implies that
solutions of this type are allowed in the region $2<q<2.13$. 
Finally, bright-dark ($bd$) vector soliton solutions ($f$, $l$) of the form:
%%%%%%
\begin{eqnarray}
f_{bd}=a_1{\rm sech}\left(\beta_{bd}x_1\right)e^{-i\nu_1t_2},
\label{fbd} \\
l_{bd}=a_{2}\tanh\left(\beta_{bd}x_1\right)e^{-i\nu_2t_2},
\label{lbd}
\end{eqnarray}
%%%%
as well as dark-bright ($db$) solutions (by interchanging $f_{bd}\leftrightarrow l_{bd}$) can 
also be found.
Solitons of the $bd$ ($db$) type are possible in the regions $0<q<0.76$, $q>2.63$ 
($1.06<q<1.88$).
In all cases the soliton widths, amplitudes and frequencies are connected through the equations
shown in  Table~\ref{table:sol}. 
%in the different regions.
%%%%%%%%%
Figure~\ref{fig1} shows the coupling constants $g_{ij}$ as functions of $q$, and the region
of existence for each different vector soliton is highlighted.
Using Eqs.~(\ref{sol1a})-(\ref{sol1h}) and the
aforementioned soliton solutions of Eqs.~(\ref{nlsf})-(\ref{nlsl})
we can obtain localized approximate solutions (to order $\epsilon$)
for the fields $A$ and $H$:
%%%%%
\begin{eqnarray}
A(x,t)&\approx & 2\epsilon f(\epsilon \beta x)\cos\left[\left(1+\epsilon ^2 \nu_1\right)t\right],
\label{appra} \\
H(x,t)&\approx & 2\epsilon l(\epsilon \beta x)\cos\left[\left(q+\epsilon ^2 \nu_2\right)t\right].
\label{apprh}
\end{eqnarray}
%%%%
Inserting the profiles of $f$, $l$ in Eqs.~(\ref{appra})-(\ref{apprh}), we obtain
different classes of approximate solutions.
Those corresponding to $bb$-solitons as in Eqs.(\ref{fbb})-(\ref{lbb}), 
will have the form of oscillons for both fields.
On the other hand $dd$-solitons correspond to solutions where both fields have the
form of oscillating kinks. Finally $bd$ ($db$) solitons will result in an oscillon for the
field $A$ ($H$) and an oscillating kink for $H$ ($A$).

%%%%%%%%%%%%%%%%
\subsection{Modulation Instability}

In this section we will explore the impact of the modulation instability mechanism to the
solution space of the Abelian-Higgs model. To this end, we examine the stability  
of plane wave solutions of the CNLS Eqs.~(\ref{nlsf})-(\ref{nlsl}).
The mechanism of modulation instability (MI)~\cite{mi} is an important property of the NLS
equation, revealing localized structures that a system supports,
e.g. ${\rm sech}-shaped$ solutions of the form of Eqs.~(\ref{fbb})-(\ref{lbb}).
This mechanism also gives information about the $\tanh-shaped$ solutions,
e.g. Eqs.~(\ref{fdd})-(\ref{ldd}), since
instability of plane waves leads to unstable background for this type of solutions.
We consider the following ansatz:
%%%%
\begin{eqnarray}
f(x_1,t_2)&=&\left(f_0+\delta f \right)e^{-i\Omega_1 t_2},
\label{planef} \\
l(x_1,t_2)&=&\left(l_0+\delta l \right)e^{-i\Omega_2 t_2},
\label{planel}
\end{eqnarray}
%%%%
where $f_0$ and $l_0$ are the amplitudes of the plane wave solutions
of the CNLS equations, and $\Omega_{1,2}$ are their frequencies satisfying
the dispersion relations
%%%%
\begin{eqnarray}
\omega_1 \Omega_1 &=&g_{11}|f_0|^2+g_{12}|l_0|^2,
\label{disf} \\
\omega_2 \Omega_2 &=&g_{21}|f_0|^2+g_{22}|l_0|^2.
\label{disl}
\end{eqnarray}
%%%%
The small amplitude perturbations i.e.
$\delta f/f, \delta l/l\ll1$, are complex functions of the form
$\delta f=u_1+iw_1$, $\delta l=u_2+iw_2$. The real functions $u_j, w_j$ are considered
to be of the general form:
%%%%%
\begin{eqnarray}
u_j&=&u_{0j}\exp [i\left(Kx_1-\Omega t_2\right)]+ c.c.,
\label{uj} \\
w_j&=&w_{0j}\exp [i\left(Kx_1-\Omega t_2\right)]+ c.c.,
\label{wj}
\end{eqnarray}
%%%%
where the amplitudes $u_{0j}$, $w_{0j}$ are constants while K is the wavenumber and $\Omega$
the frequency of the perturbation. Substituting Eqs.~(\ref{uj})-(\ref{wj})
to the CNLS Eqs.~(\ref{nlsf})-(\ref{nlsl}) leads to an algebraic system of equations, the
determinant of which has to be zero. This compatibility
condition leads to the following equation:
\begin{eqnarray}
A \Omega^4-\left(\omega^2_1 B_2+\omega^2_2 B_1\right)\Omega^2 +\left(B_1B_2-\Gamma
\right)=0,
\label{tri}
\end{eqnarray}
%%%%
where $A=\left(\omega_1\omega_2\right)^2$, $B_j=K^2/2\left(K^2/2+2g_{jj}f^2_0\right)$
with $j=1,2$ and $\Gamma=K^4g_{12}g_{21}(f_0 l_0)^2$. Requiring real roots
of Eq.~(\ref{tri}) we are led to the following stability conditions:
%%%%
\begin{eqnarray}
g_{11}g_{22}-g_{12}g_{21}>0,  \quad g_{11}\geq 0, \quad & g_{22}\geq 0.
\label{ineq}
\end{eqnarray}
%%%%
As seen from Eq.~(\ref{gij}), there are no real values of the parameter $q$ satisfying
the last inequality of Eq.~(\ref{ineq}). Thus plane wave solutions are unstable
under small perturbations.
The aforementioned result implies that solutions including $d$ components,
although they are allowed by the CNLS system in Eqs.~(\ref{nlsf})-(\ref{nlsl}), 
are expected to be unstable due to the modulation instability.
This can be seen from the following fact: the dark-component
in the solitons of  Eqs.~(\ref{fdd})-(\ref{lbd}), takes asymptotically  
the form $\sim \pm a_i\exp[-i\nu_i t]$ at $x\rightarrow\pm\infty$.
Hence away from its core,
the solution is a plane wave and any perturbation will lead to the appearance of
MI and the generation of oscillons.
Thus we argue that localized solutions in the form of kinks, are not
supported in the setting where $A$ and $H$ are of the same order.
%%%%%
\begin{figure}
%\centering
%\hspace{-80pt}
%\begin{subfigure}[b]
\includegraphics[scale=0.36]{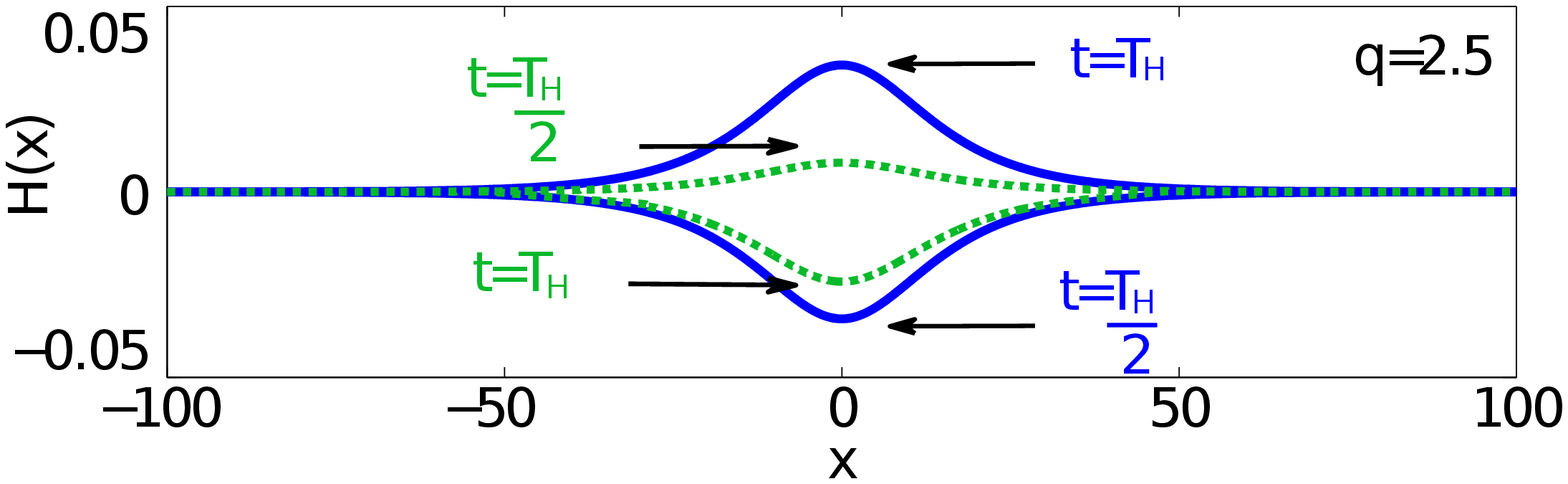}
%\caption*{\hspace{30pt}(a)}
%\end{subfigure}
%\begin{subfigure}{0.5}
\includegraphics[scale=0.4]{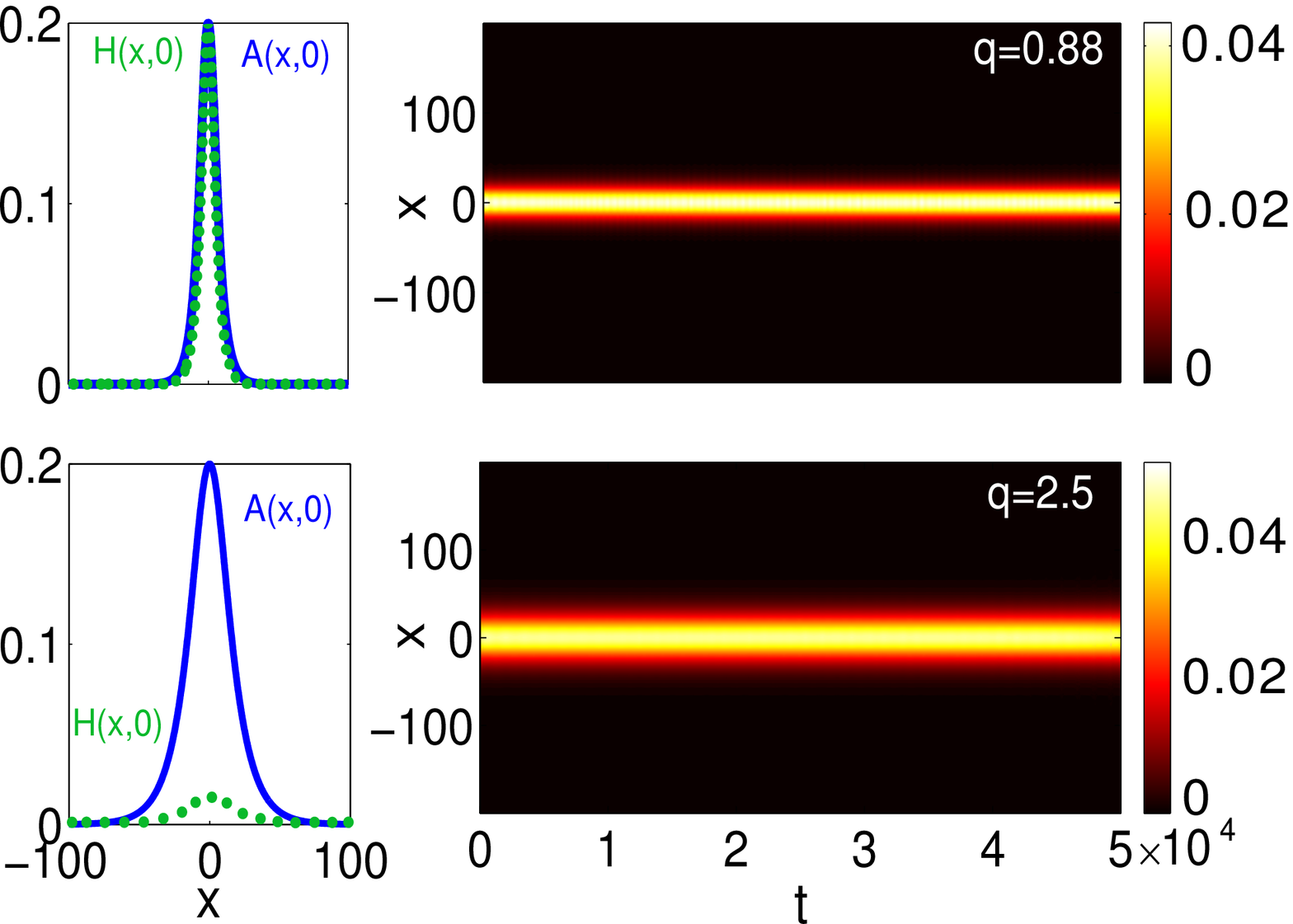}
%\caption*{\hspace{80pt}(b)}
%\end{subfigure}
\caption{(Color online): Top panel shows a plot of the Higgs field for 
$t=T_H$, $t=T_H/2$ and $q=2.5$. Solid (blue) lines refer to the symmetric oscillations of the 
field $H(x)$. For comparison, in the same figure dotted (green) lines depict the oscillon in the 
asymmetric case. Middle-Bottom: left panels depict the profiles of the two fields $A$ [solid 
(blue) lines], and $H$ [dotted (green) line], for $t=0$, corresponding to oscillon solutions 
of Eqs.~(\ref{appra})-(\ref{apprh}) for different values of $q$.  
Right panels: contour plots showing the evolution of the energy density
$E(x,t)$ for initial conditions corresponding to the left panel. The total time of integration
is $t=5\times10^4$ and the different values of $q$ are depicted in the upper right corner
of each contour.}
\label{fig2}
\end{figure}
%%%%%

According to the above analysis, the relevant solutions
for the functions $f(x_1,t_2)$ and $l(x_1,t_2)$ are the $bb$ solitons, which are
found in the three regions defined in the previous paragraph by Eqs.~(\ref{fbb})-(\ref{lbb})
(see also shaded areas of Fig.~\ref{fig1}). The profile of a Higgs field oscillon   
with $q=2.5$, is shown in the top panel Fig.~\ref{fig2} with a solid (blue) line, for a half and 
a total period of oscillation. Notice that in contrast to previous findings~\cite{prd},  
the Higgs field performs symmetric oscillations around the ``classical vacuum''(i.e $H=0$).
For comparison a typical solution of an asymmetric oscillon of Ref.~\cite{prd} is also shown in 
Fig.~\ref{fig2} [dotted (green) lines] for the same value of $q$. As already discussed, the two 
types of oscillons (asymmetric, symmetric) describe different scenarios for the underlying 
physics: the asymmetric solutions are valid in the case of weakly broken symmetry, i.e. just 
beyond the associated critical point, while the symmetric solutions occur when the gauge 
symmetry is strongly broken, i.e. far beyond the critical point. 

It is also important to note that in the region $1.88<q<2$, the amplitudes of the
second order expansions of the fields [cf. Eqs.~(\ref{a2})-(\ref{h2})]
become larger than $\mathcal{O}(1)$ and the perturbation scheme
collapses.
Thus, in this case oscillon solutions are not expected to exist for the exact system.   
However, our numerical results show that an initial condition corresponding
to $bb$ solitons in this region also leads to robust oscillon solutions.

\section{Numerical results: oscillon's longevity}

According to the analysis of the previous section, robust localized solutions
of the original system of equations~(\ref{eqa})-(\ref{eqh}) in the form of NLS bright-bright 
solitons, are expected to be found in the two parameter regions
$0.76<q<1.06$ (RI) and $2.13<q<2.63$ (RII) [cf. Fig.~\ref{fig1}].
To clarify this issue we perform numerical integration of the exact system
Eqs.~(\ref{eqa})-(\ref{eqh}) for a wide range of $q$ values, using as initial conditions
the approximate solutions Eqs.~(\ref{appra})-(\ref{apprh}) for the case of bb-solitons. 
Our main interest is to confirm the existence of these structures and explore their 
long time dynamics.
We use a fourth-order Runge-Kutta integrator for the time propagation with a lattice of length
$L=400$, lattice spacing $dx=0.1$, time step $dt=10^{-2}$ and $\epsilon=0.1$. With
this choice of parameters the numerical integration conserves the energy for the total time
interval of our simulations up to the order of $10^{-3}$.

In the middle panel of Fig.~\ref{fig2} we show an example of the evolution of an oscillon
in RI for $q=0.88$. The middle left panel depicts the oscillon profile at $t=0$, 
were solid (blue) line and dotted (green)line present $A$ and $H$ respectively.
The evolution of the energy density $E(x,t)$ of Eq.~\ref{energy}, is shown in the middle right 
panel for a time interval of $t=5\times 10^4$, corresponding to $\sim10^4$ oscillations for both
fields. The energy density remains localized throughout the simulation, indicating the
robustness of the oscillons in RI. We have confirmed (results
not shown here) the existence and longevity of oscillons in this region for different
values of $q$.

Next, an example of an oscillon in RII and for $q=2.5$
is shown in the bottom panels of Fig.~\ref{fig2}. The evolution of the
energy density in the right panel again confirms the existence of the respective oscillon
solution and illustrates its longevity, since it turns out to be robust after performing
at least $\sim10^4$ oscillations. Note that for this example and 
for $q \in [2.3, 2.63]$ (RII), the amplitude of $H$ gets suppressed
so that the ratio of the amplitudes of the fields $H/A$ is close to the value of 
the perturbation parameter $\epsilon$. In this sense for this region of parameter values
one recovers the scenario of small Higgs amplitude, explored in our previous work~\cite{prd}.
However the solutions presented here have a different profile than those in~\cite{prd}.
In particular, the Higgs field in~\cite{prd} exhibits asymmetric oscillations
with respect to the origin (i.e. $H=0$), while the oscillon of Eq.~(\ref{apprh}) is symmetric 
[see also top panel of Fig.~\ref{fig2}].
%%%%%
\begin{figure}[tbp]
%\hspace{-25pt}
\centering
\includegraphics[scale=0.4]{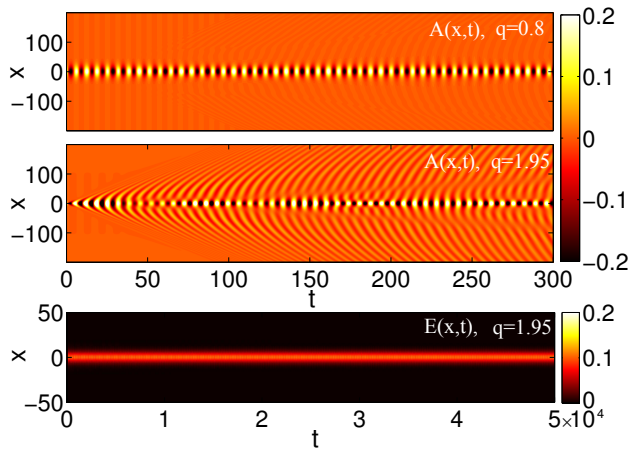}
\caption{(Color online): Top and middle panels: Contour plots showing the evolution of the gauge
field $A(x,t)$ for different values of $q$, as indicated in the upper right corner of each.
Bottom: Contour of the energy density $E(x,t)$ for a time interval $t=5\times10^4$.}
\label{fig3}
\end{figure}
%%%%%%%%%%%%%%%%
Formally, this is due to the fact that the Higgs field in~\cite{prd} obeys a linear equation
with an external source generated by the gauge field, while here the Higgs field obeys an
NLS equation coupled with the gauge field. As previously explained, these two types of oscillons 
correspond to different scenarios for the underlying physics.
We have performed simulations and confirmed the existence of
robust oscillons described by Eqs.~(\ref{fbb})-(\ref{lbb}) in both
RI and RII for various values of $q$.

Additionally, our numerical findings support the existence of oscillons in the region where the
perturbation scheme is not valid, i.e. for $1.88<q<2$.
Using the same initial condition as above, we have integrated Eqs.~(\ref{eqa})-(\ref{eqh}) for
$q=1.95$ and the results are shown in the middle and bottom panels of Fig.~\ref{fig3}.
Since the system does not appear to support oscillons of the form of 
Eqs.~(\ref{fbb})-(\ref{lbb}), it is natural to expect a distortion of these type of solutions.
Indeed, a large amount of radiation is emitted
from the vicinity of the initially localized structure, as shown in the middle panel of
Fig.~\ref{fig3}. 
This panel depicts the short time evolution of $A$, and it is clearly seen that the distortion 
of the core starts very early.
However, after sufficient time, a localized oscillating structure,
different from  the initial one, is formed 
and remains undistorted through the time of the evolution.
The energy density of this oscillon is shown in the  
bottom panel of Fig.~\ref{fig3}. 
The same qualitative result was observed for different values of
$q$ in this region (results not shown here), suggesting that oscillons do exist 
but they are not described by Eqs.~(\ref{fbb})-(\ref{lbb}). 
For completeness, the short time evolution of an oscillon in RI 
is plotted in the top panel of Fig.~\ref{fig3}, to highlight the fact
that the radiation  of the {\it true} oscillon solution (see Ref.~\cite{segur}) has much
smaller amplitude than the oscillon. 

Next, we perform numerical integration of the exact system of equations in the
regions of $q$ where solutions in the form of Eqs.~(\ref{fbd})-(\ref{lbd}) are 
expected to be subject to the MI mechanism. 
%%%%%
\begin{figure}[tbp]
\centering
\includegraphics[scale=0.35]{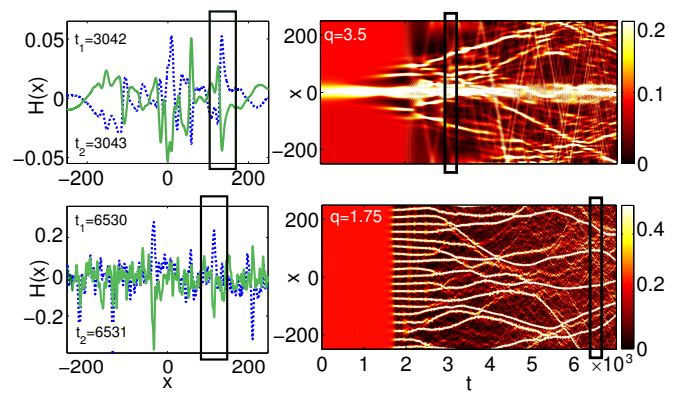}
\caption{(Color online): Top left panel: profile snapshots of the field $H$ at
two different instants, $t_1=3042$ dashed (blue) line and $t_2=3043$ solid (green) line. 
Solid (black) box focuses on a single oscillon showing
its oscillation in half period.   
Top right panel:
Contour plot showing the evolution of the energy density
$E(x,t)$ for a $bd$ initial condition. The black
box indicates the time of the snapshot. Bottom panels are the same as the 
top panels for plane wave initial conditions.
In both contours the values of $q$ are indicated in the upper left corner,
and the time interval is $t=7\times10^3$.}
\label{fig4}
\end{figure}
%%%%%%%%%%%%%%%%
The evolution of the energy density for such a $bd$ soliton is shown in the top right panel of 
Fig.~\ref{fig4} for $q=3.5$. We observe that at $t\sim 2\times 10^3$ 
the initially localized solution deforms and the instability settles in. Through the
instability, localized structures occur on top of the initial solution (see e.g. the black
box around $t=3042$) having the form of oscillons. As an example, the profile of the Higgs
field for $t_1=3042$ [dashed (blue) line] and $t_2=3043$ [solid (green) line] is 
shown in the top left panel of Fig.~\ref{fig4}. The
solid (black) box indicates an oscillon performing a half oscillation period. The above result 
is in agreement with our analytical findings regarding the instability of the oscillating kinks.

We complete our numerical analysis by showing in the bottom panels
of Fig.~\ref{fig4} the generation of oscillons at $q=1.75$, that is in the $db$ region.  
We have used initial conditions of the form:
$A=H=C_0\left[1+\delta \cos(Kx)\right]$, where $C_0=0.05$ and $\delta=10^{-2}$
are the plane wave and the perturbation amplitude respectively and $K=0.025$ 
is a wavenumber inside the instability band given by Eq.~(\ref{tri}). 
The initial, almost flat profile of the energy density deforms into a periodic pattern at
$t\approx2\times 10^3$, due to the modulational-instability-induced exponential growth 
of the wavenumber $K$. At later times moving oscillons are formed, which are subject to 
collisions and eventually some survive and some annihilate.
The time interval indicated by the solid (black) box in the bottom right panel of 
Fig.~\ref{fig4} contains two time instants for which we show the Higgs field profile in the 
respective left panel.
The solid (black) box in the left panel, focuses on a single oscillon at $x\approx100$ 
performing a half oscillation period.
Dashed (blue) line corresponds to $t_1=6530$ and solid (green) line to $t_2=6531$. 
The oscillon performs oscillations with period $T_{osc}=2\pi/q$ in agreement with the 
analytical predictions.  

\section{Conclusions and discussion}

In conclusion, we have presented approximate analytical solutions of the $(1+1)$-dim Abelian-
Higgs model, when the amplitudes of the gauge and the Higgs field are of the same order. 
Employing a multiple scale perturbation theory we reduced the original set of equations into a 
system of CNLS equations which admits different types of 
exact analytical solutions, depending on the  parameter $q$ (i.e. the ratio of two field 
masses).

Our analysis reveals that bright-bright solitons of the CNLS equations lead to robust long-lived 
oscillon solutions of the original system. These oscillons are characterized by symmetric 
oscillations around the ``classical vacuum", for both the gauge and the Higgs field, describing 
excitations which may occur when the gauge symmetry is strongly broken (in contrast to the 
previous results of~\cite{prd}).
 
Direct numerical simulations of the original system of equations of motion,
confirm the robustness of the obtained solutions for times up to $5\times 10^4$
oscillation periods. These solutions are shown to exist
in two different regions:  $0.76<q<1.06$ (RI) and $2.13<q<2.63$ (RII). 
In the context of superconductor phenomenology a possible interpretation of our solutions is 
that oscillons with mass ratio less than unity (RI) correspond to type-I superconductors, 
while oscillons of RII, to type-II superconductors. This analogy is compatible with the 
statements reported in Ref.~\cite{gleiser21}, for the Abelian-Higgs in $(3+1)$-dim.       

Using the derived CNLS system we have also shown that for any value of the
parameter $q$, plane waves are subject to the modulation instability mechanism.
Dark-bright (bright-dark) solitons and the corresponding oscillating kinks
were found to be unstable, and their instability was shown to lead into oscillon-like
patterns. 
 
The methodology used in the present work, may be applied in order to obtain
novel localized structures in different settings, either with different potentials
for the scalar field, or in the presence of non-Abelian gauge fields~\cite{farhi05} and/or even 
in higher dimensional settings. Finally, based on the obtained solutions it would be interesting 
to further explore oscillon-oscillon interactions and understand their impact on the long time 
field dynamics.

\end{document}